\documentclass{biometrika}

\usepackage{amsmath}
\usepackage{latexsym, amssymb, graphicx, mathrsfs}

\usepackage{times}
\usepackage{bm}
\usepackage{natbib}

\usepackage[plain,noend]{algorithm2e}

\makeatletter
\renewcommand{\algocf@captiontext}[2]{#1\algocf@typo. \AlCapFnt{}#2} 
\def\@algocf@capt@plain{top}
\renewcommand{\algocf@makecaption}[2]{%
  \addtolength{\hsize}{\algomargin}%
  \sbox\@tempboxa{\algocf@captiontext{#1}{#2}}%
  \ifdim\wd\@tempboxa >\hsize
    \hskip .5\algomargin%
    \parbox[t]{\hsize}{\algocf@captiontext{#1}{#2}}
  \else%
    \global\@minipagefalse%
    \hbox to\hsize{\box\@tempboxa}
  \fi%
  \addtolength{\hsize}{-\algomargin}%
}
\makeatother

\def\Bka{{\it Biometrika}}

\newcommand{\ba}{a}
\newcommand{\bb}{b}

\newcommand{\bx}{x}
\newcommand{\by}{y}

\newcommand{\bI}{I}

\newcommand{\bX}{X}

\newcommand{\bone}{1}
\newcommand{\bzero}{0}
\newcommand{\bveps}{\varepsilon}

\newcommand{\bbeta}{\beta}
\newcommand{\bdelta}{\delta}

\newcommand{\bgamma}{\gamma}

\newcommand{\hbbeta}{\widehat\bbeta}

\newcommand{\hbeta}{\widehat\beta}
\newcommand{\tbbeta}{\widetilde\bbeta}

\newcommand{\FS}{\mathrm{FS}}

\newcommand{\veps}{\varepsilon}

\newcommand{\sgn}{\mathrm{sgn}}
\newcommand{\supp}{\mathrm{supp}}

\def\t{^T}

\begin{document}


\jname{}
\jyear{}
\jvol{}
\jnum{}
\copyrightinfo{}


\markboth{Y. Fan \and J. Lv}{Combined $L_1$ and concave regularization}

\title{Asymptotic properties for combined $L_1$ and concave regularization}

\author{Yingying Fan \and Jinchi Lv}
\affil{Data Sciences
and Operations Department, 
University of Southern California, Los Angeles, California 90089, U.S.A. \email{fanyingy@marshall.usc.edu} \email{jinchilv@marshall.usc.edu}}

\maketitle

\begin{abstract}
Two important goals of high-dimensional modeling are prediction and variable selection. In this article, we consider regularization with combined $L_1$ and concave penalties, and study the sampling properties of the global optimum of the suggested method in ultra-high dimensional settings. The $L_1$-penalty provides the minimum regularization needed for removing noise variables in order to achieve oracle prediction risk, while concave penalty imposes additional regularization to control model sparsity. In the linear model setting, we prove that the global optimum of our method enjoys the same oracle inequalities as the lasso estimator and admits an explicit bound on the false sign rate, which can be asymptotically vanishing. Moreover, we establish oracle risk inequalities for the method and the sampling properties of computable solutions. Numerical studies suggest that our method yields more stable estimates than using a concave penalty alone.
\end{abstract}


\section{Introduction} \label{Sec1}
Prediction and variable selection are two important goals in many contemporary large-scale problems. Many regularization methods in the context of penalized empirical risk minimization have been proposed to select important covariates. See, for example, Fan \& Lv (2010) for a review of some recent developments in high-dimensional variable selection. Penalized empirical risk minimization has two components: empirical risk for a chosen loss function for prediction, and a penalty function on the magnitude of parameters for reducing model complexity. The loss function is often chosen to be convex. The inclusion of the regularization term helps prevent overfitting when the number of covariates $p$ is comparable to or exceeds the number of observations $n$.

Generally speaking, two classes of penalty functions have been proposed in the literature: convex ones and concave ones. When a convex penalty such as the lasso penalty \citep{Tibshirani96} is used, the resulting estimator is a well-defined global optimizer. For the properties of $L_1$-regularization methods, see, for example, \cite{CDS99}, \cite{EHJT04}, \cite{Zou06}, \cite{CT07}, \cite{RZ07}, and \cite{BRT09}. In particular, \cite{BRT09} proved that using the $L_1$-penalty leads to estimators satisfying the oracle inequalities under the prediction loss and $L_q$-loss, with $1 \leq q \leq 2$, in high-dimensional nonparametric regression models. An oracle inequality means that with an overwhelming probability, the loss of the regularized estimator is within a logarithmic factor, a power of $\log p$, of that of the oracle estimator, with the power depending on the chosen estimation loss. Despite these nice properties, the $L_1$-penalty tends to yield a larger model than the true one for optimizing predictions, and many of the selected variables may be insignificant, showing that the resulting method may not be ideal for variable selection. The relatively large model size also reduces the interpretability of the selected model.

Concave penalties, on the other hand, have been shown to lead to nice variable selection properties. The oracle property was introduced in \cite{FL01} to characterize the performance of concave regularization methods, in relation to the oracle procedure knowing the true sparse model in advance. In fixed dimensions, concave regularization has been shown to have the oracle property, recovering the true model with asymptotic probability one. This work has been extended to higher dimensions in different contexts, and the key message is the same. See, for example, \cite{LF09}, \cite{Zhang10}, and \cite{FLv11}. In particular, the weak oracle property, a surrogate of the oracle property, was introduced in \cite{LF09}. When $p > n$, it is generally difficult to study the properties of the global optimizer for concave regularization methods. Thus, most studies have focused on some local optimizer that has appealing properties in high-dimensional settings. The sampling properties of the global optimizers for these methods are less well-understood in high dimensions.

In this article, we characterize theoretically the global optimizer of the regularization method with the combined $L_1$ and concave penalty, in the setting of the high-dimensional linear model. We prove that the resulting estimator combines the prediction power of the $L_1$-penalty and the variable selection power of the concave penalty. On the practical side, the $L_1$-penalty contributes the minimum amount of regularization necessary to remove noise variables for achieving oracle prediction risk, while the concave penalty incorporates additional regularization to control model sparsity. On the theoretical side, the use of an $L_1$-penalty helps us to study the various properties of the global optimizer. Specifically, we prove that the global optimizer enjoys the oracle inequalities under the prediction loss and $L_q$-loss, with $1 \leq q \leq 2$, as well as an asymptotically vanishing bound on false sign rate. We also establish its oracle risk inequalities under various losses, as well as the sampling properties of computable solutions. In addition, we show that the refitted least-squares estimator can enjoy the oracle property, in the context of \cite{FL01}. These results are also closely related to those in \cite{ZZ12}. Our work complements theirs in three important respects. First, the bound on the number of false positives in \cite{ZZ12} is generally of the same order as the true model size, while our bound on the stronger measure of the rate of false signs can be asymptotically vanishing. Second, our estimation and prediction bounds depend only on the universal regularization parameter for the $L_1$-component and are free of the regularization parameter $\lambda$ for the concave component, whereas the bounds in \cite{ZZ12} generally depend on $\lambda$ alone. Third, our oracle risk inequalities are new and stronger than those for losses, since the risks involve the expectations of losses and thus provide a more complete view of the stability of the method. It is unclear whether the concave method alone may enjoy similar risk bounds.

Our proposal shares a similar spirit to that in \cite{LW07}, who proposed a combination of $L_0$- and $L_1$-penalties for variable selection and studied its properties in linear regression with fixed dimensionality. Their new penalty yields more stable variable selection results than the $L_0$-penalty, and outperforms both $L_0$- and $L_1$-penalties in terms of variable selection, while maintaining good prediction accuracy. Our theoretical results and numerical study reveal that this advantage still exists in high dimensions and for more general concave penalties. Our work differs from theirs in two main respects: we provide more complete and unified theory in ultra-high dimensional settings, and we consider a large class of concave penalties with only mild conditions on their shape. The idea of combining strengths of different penalties has also been exploited in, for example, \cite{ZZ09}.

\section{Model setting} \label{Sec2}

Consider the linear regression model
\begin{equation} \label{e001}
\by = \bX \bbeta + \bveps,
\end{equation}
where $\by = (Y_1, \ldots, Y_n)\t$ is an $n$-dimensional vector of responses, $\bX = (\bx_1, \ldots, \bx_p)$ is an $n \times p$ design matrix, $\bbeta = (\beta_1, \ldots, \beta_p)\t$ is an unknown $p$-dimensional vector of regression coefficients, and $\bveps = (\veps_1, \ldots, \veps_n)\t$ is an $n$-dimensional vector of noises. We are interested in variable selection when the true regression coefficient vector $\bbeta_0 = (\beta_{0,1}, \ldots, \beta_{0,p})\t$ has many zero components. The main goal is to effectively identify the true underlying sparse model, that is, the support $\supp(\bbeta_0) = \{j = 1, \ldots, p: \beta_{0, j} \neq 0\}$, with asymptotic probability one, and to efficiently estimate the nonzero regression coefficients $\beta_{0,j}$'s. A popular approach to estimating sparse $\bbeta_0$ is penalized least squares, which regularizes the conventional least-squares estimation by penalizing the magnitude of parameters $|\beta_j|$. A zero component of the resulting estimate indicates that the corresponding covariate $\bx_j$ is screened from the model.

Penalized least-squares estimation minimizes the objective function
\[ (2n)^{-1} \|\by - \bX \bbeta\|_2^2 + \|p_\lambda(\bbeta)\|_1 \]
over $\bbeta \in\mathbb{R}^p$, where we use the compact notation $p_\lambda(\bbeta) = p_\lambda(|\bbeta|) = (p_\lambda(|\beta_1|), \ldots, p_\lambda(|\beta_p|))\t$ with $|\bbeta| = (|\beta_1|, \ldots, |\beta_p|)\t$, and $p_\lambda(t)$, $t \in [0, \infty)$, is a penalty function indexed by the regularization parameter $\lambda \geq 0$. The lasso (Tibshirani, 1996) corresponds to the $L_1$-penalty $p_{\lambda}(t) = \lambda t$. As shown in Bickel et al. (2009), the lasso enjoys the oracle inequalities for prediction and estimation, but it tends to yield large models. Concave penalties have received much attention due to their oracle properties. Yet, as discussed in \S\ref{Sec1}, the sampling properties of the global optimizer for concave regularization methods are relatively less well-understood in high dimensions. To overcome these difficulties, we suggest combining the $L_1$-penalty $\lambda_0 t$ with a concave penalty $p_\lambda(t)$, and study the resulting regularization problem
\begin{equation} \label{e006}
\min_{\bbeta\in \mathbb{R}^p} \Big\{(2n)^{-1} \|\by - \bX \bbeta\|_2^2 + \lambda_0\|\bbeta\|_1 + \|p_\lambda(\bbeta)\|_1\Big\},
\end{equation}
where $\lambda_0 = c \{(\log p)/n\}^{1/2}$ for some positive constant $c$. Throughout the paper, we fix such a choice of the universal regularization parameter for the $L_1$-penalty, and the minimizer of (\ref{e006}) is implicitly referred to as the global minimizer. The $L_1$-component $\lambda_0 \|\bbeta\|_1$ helps study the global minimizer of (\ref{e006}), and reflects the minimum amount of regularization for removing the noise in prediction. The concave component $\|p_\lambda(\bbeta)\|_1$ serves to adapt the model sparsity for variable selection.

\section{Main results} \label{Sec3}

\subsection{Hard-thresholding property} \label{Sec3.1}
To understand why the combination of $L_1$- and concave penalties can yield better variable selection than can the $L_1$-penalty alone, we consider the hard-thresholding penalty
$p_{\text{H}, \lambda}(t) = 2^{-1}\{\lambda^2 - (\lambda - t)_+^2\}$, $t \geq 0$. Assume that each covariate $\bx_j$ is rescaled to have $L_2$-norm $n^{1/2}$. Let $\hbbeta = (\hbeta_1, \ldots, \hbeta_p)\t$ be the global minimizer of (\ref{e006}) with $p_{\lambda}(t) = p_{\text{H}, \lambda}(t)$. The global optimality of $\hbbeta$ entails that each $\hbeta_j$ is the global minimizer of the corresponding univariate penalized least-squares problem along the $j$th coordinate. All these univariate problems share a common form, with generally different scalar $z$'s,
\begin{equation} \label{e010}
\hbeta(z) = \mbox{argmin}_{\beta \in \mathbb{R}} \left\{2^{-1} (z - \beta)^2 + \lambda_0 |\beta| + p_{\text{H}, \lambda}(|\beta|)\right\},
\end{equation}
since all covariates have $L_2$-norm $n^{1/2}$. Simple calculus shows that the solution in (\ref{e010}) is
\begin{equation} \label{e013}
\hbeta(z) = \sgn(z) (|z| - \lambda_0) 1_{\{|z| > \lambda + \lambda_0\}},
\end{equation}
so the resulting estimator has the same feature as the hard-thresholded estimator: each component is either zero or of magnitude larger than $\lambda$. This provides an appealing distinction between insignificant covariates, whose coefficients are zero and should be estimated as such, and significant covariates, whose coefficients are significantly nonzero and should be estimated as nonzero, improving the variable selection performance of soft-thresholding by $L_1$-penalty.

The hard-thresholding feature is shared by many other penalty functions, as now shown.

\begin{proposition} \label{Prop1}
Assume that $p_{\lambda}(t)$, $t \geq 0$, is increasing and concave with $p_\lambda(t) \geq p_{\text{H}, \lambda}(t)$ on $[0, \lambda]$, $p_\lambda'\{(1 - c_1) \lambda\} \leq c_1 \lambda$ for some $c_1 \in [0, 1)$, and $-p''_\lambda(t)$ decreasing on $[0, (1 - c_1) \lambda]$. Then any local minimizer of (\ref{e006}) that is a global minimizer in each coordinate has the hard-thresholding feature that each component is either zero or of magnitude larger than $(1 - c_1) \lambda$.
\end{proposition}

Although we used the derivatives $p'_{\lambda}(t)$ and $p''_{\lambda}(t)$ in the above proposition, the results continue to hold if we replace $-p'_{\lambda}(t)$ with the subdifferential of $-p_{\lambda}(t)$, and $-p_{\lambda}''(t)$ with the local concavity of $p_{\lambda}(t)$ at point $t$, when the penalty function is nondifferentiable at $t$ (Lv \& Fan, 2009). The hard-thresholding penalty $p_{\text{H}, \lambda}(t)$ satisfies conditions of Proposition \ref{Prop1}, with $c_1 = 0$. This class of penalty functions also includes, for example, the $L_0$-penalty and the smooth integration of counting and absolute deviation penalty (Lv \& Fan, 2009), with suitably chosen $c_1 \in [0, 1)$ and tuning parameters.

\subsection{Technical conditions} \label{Sec3.2}

We consider a wide range of error distributions for the linear model (\ref{e001}). Throughout this paper, we make the following assumption on the distribution of model error $\bveps$:
\begin{equation} \label{e024}
{\rm pr}(\|n^{-1} \bX\t \bveps\|_\infty > \lambda_0/2) = O(p^{-c_0}),
\end{equation}
where $c_0$ is some arbitrarily large, positive constant depending only on $c$, the constant defining $\lambda_0$. This condition was imposed in \cite{FLv11}, who showed for independent $\veps_1, \ldots, \veps_n$ that Gaussian errors and bounded errors satisfy (\ref{e024}) without any extra assumption, and that light-tailed error distributions satisfy (\ref{e024}) with additional mild assumptions on the design matrix $\bX$.

Without loss of generality, we assume that only the first $s$ components of $\bbeta_0$ are nonzero, where the true model size $s$ can diverge with the sample size $n$. Write the true regression coefficient vector as $\bbeta_0 = (\widetilde{\bbeta}_{0,1}\t, \widetilde{\bbeta}_{0,2}\t)\t$ with $\widetilde{\bbeta}_{0,1} = (\beta_{0, 1}, \ldots, \beta_{0, s})\t \in \mathbb{R}^{s}$ the subvector of all nonzero coefficients and $\widetilde{\bbeta}_{0,2} = \bzero$, and let $p_\lambda(\infty) = \lim_{t \rightarrow \infty} p_\lambda(t)$.
We impose the following conditions on the design matrix and penalty function, respectively.

\begin{condition} \label{cond1}
For some positive constant $\kappa_0$, $\min_{\|\bdelta\|_2 = 1,\ \|\bdelta\|_0 < 2 s}n^{-1/2} \|\bX \bdelta\|_2 \geq \kappa_0$ and
\begin{equation}\label{002}
\kappa = \kappa(s, 7) = \min_{ \bdelta \neq\bzero, \ \|\widetilde{\bdelta}_{2}\|_1 \leq 7 \|\widetilde{\bdelta}_{1}\|_1} \big\{n^{-1/2} \|\bX \bdelta\|_2/(\|\widetilde{\bdelta}_{1}\|_2 \vee \|\widetilde{\bdelta}'_2\|_2)\big\} > 0,
\end{equation}
where $\bdelta = (\widetilde{\bdelta}_1\t, \widetilde{\bdelta}_2\t)\t$ with $\widetilde{\bdelta}_1 \in \mathbb{R}^s$ and $\widetilde{\bdelta}'_2$ the subvector of $\widetilde{\bdelta}_2$ consisting of the components with the $s$ largest absolute values.
\end{condition}

\begin{condition} \label{cond2}
The penalty $p_{\lambda}(t)$ satisfies the conditions of Proposition \ref{Prop1} with $p_\lambda'\{(1 - c_1) \lambda\} \leq \lambda_0/4$, and $\min_{j = 1, \ldots, s} |\beta_{0, j}| > \max\{(1 - c_1) \lambda, 2 \kappa_0^{-1} p_\lambda^{1/2}(\infty)\}$.
\end{condition}

The first part of Condition \ref{cond1} is a mild sparse eigenvalue condition, and the second part combines the restricted eigenvalue assumptions in \cite{BRT09}, which were introduced for studying the oracle inequalities for the lasso estimator and Dantzig selector \citep{CT07}. To see the intuition for (\ref{002}), recall that the ordinary least-squares estimation requires that the Gram matrix $\bX\t\bX$ be positive definite, that is,
\begin{equation}\label{001}
\min_{\bzero\neq\bdelta \in \mathbb{R}^p}\big\{ n^{-1/2}\|\bX\bdelta\|_2/ \|\bdelta\|_2 \big\}>0.
\end{equation}
In the high-dimensional setting $p>n$, condition (\ref{001}) is always violated. Condition \ref{cond1} replaces the norm $\|\bdelta\|_2$ in the denominator of (\ref{001}) with the $L_2$-norm of only a subvector of $\bdelta$. Condition \ref{cond1} also has an additional bound involving $\|\widetilde{\bdelta}_2'\|_2$. This is needed only when dealing with the $L_q$-loss with $q \in (1, 2]$. For other losses, the bound can be relaxed to
\[
\kappa = \kappa(s, 7) = \min_{ \bdelta  \neq\bzero, \ \|\widetilde{\bdelta}_{2}\|_1 \leq 7 \|\widetilde{\bdelta}_{1}\|_1} \left\{n^{-1/2} \|\bX \bdelta\|_2/\|\widetilde{\bdelta}_{1}\|_2 \right\} > 0.
\]
For simplicity, we use the same notation $\kappa$ in these bounds.

In view of the basic constraint (\ref{e012}), the restricted eigenvalue assumptions in (\ref{002}) can be weakened to other conditions such as the compatibility factor or the cone invertibility factor \citep{ZZ12}. We adopt the assumptions in \cite{BRT09} to simplify our presentation.

Condition \ref{cond2} ensures that the concave penalty $p_{\lambda}(t)$ satisfies the hard-thresholding property, requires that its tail should be relatively slowly growing, and puts a constraint on the minimum signal strength.

\subsection{Asymptotic properties of global optimum} \label{Sec3.3}
In this section, we study the sampling properties of the global minimizer $\hbbeta$ of (\ref{e006}) with $p$ implicitly understood as $\max(n, p)$ in all bounds. To evaluate the variable selection performance, we consider the number of falsely discovered signs
\[
\mbox{FS}(\hbbeta) =|\{j = 1, \ldots, p: \sgn(\hbeta_j) \neq \sgn(\beta_{0,j})\}|,
\]
which is a stronger measure than the total number of false positives and false negatives.

\begin{theorem} \label{Thm1}
Assume that Conditions \ref{cond1}--\ref{cond2} and deviation probability bound (\ref{e024}) hold, and that $p_\lambda(t)$ is continuously differentiable. Then the global minimizer $\hbbeta$ of (\ref{e006}) has the hard-thresholding property stated in Proposition \ref{Prop1}, and with probability $1 - O(p^{-c_0})$, satisfies simultaneously that
\begin{align}
\label{e034}
n^{-1/2} \|\bX (\hbbeta - \bbeta_0)\|_2 & = O(\kappa^{-1} \lambda_0 s^{1/2}),\\
\label{e035}
\|\hbbeta - \bbeta_0\|_q & = O(\kappa^{-2} \lambda_0 s^{1/q}), \quad q \in [1, 2], \\
\label{e033}
\FS(\hbbeta) & = O\{\kappa^{-4} (\lambda_0/\lambda)^2 s\}.
\end{align}
If in addition $\lambda \geq 56 (1- c_1)^{-1} \kappa^{-2} \lambda_0 s^{1/2}$, then with probability $1 - O(p^{-c_0})$, it also holds that
$\sgn(\hbbeta) = \sgn(\bbeta_0)$ and $\|\hbbeta - \bbeta_0\|_\infty = O\{\lambda_0 \|(n^{-1} \bX_{1}\t \bX_{1})^{-1}\|_\infty\}$, where $\bX_{1}$ is the $n \times s$ submatrix of $\bX$ corresponding to $s$ nonzero $\beta_{0, j}$'s.
\end{theorem}

From Theorem \ref{Thm1}, we see that if $\lambda$ is chosen such that $\lambda_0/\lambda\rightarrow 0$, then the number of falsely discovered signs $\FS(\hbbeta)$ is of order $o(s)$ and thus the false sign rate $\FS(\hbbeta)/s$ is asymptotically vanishing. In contrast, \cite{BRT09} showed that under the restricted eigenvalue assumptions, the lasso estimator, with the $L_1$-component $\lambda_0 \|\bbeta\|_1$ alone, generally gives a sparse model with size of order $O(\phi_{\max} s)$, where $\phi_{\max}$ is the largest eigenvalue of the Gram matrix $n^{-1} \bX\t \bX$. This entails that the false sign rate for the lasso estimator can be of order $O(\phi_{\max})$, which does not vanish asymptotically. Similarly, \cite{ZZ12} proved that the number of false positives of the concave regularized estimator is generally of order $O(s)$, which means that the false sign rate can be asymptotically nonvanishing.

The convergence rates in oracle inequalities (\ref{e034})--(\ref{e035}), involving both sample size $n$ and dimensionality $p$, are the same as those for the $L_1$-component alone in \cite{BRT09}, and are consistent with those for the concave component alone in \cite{ZZ12}. A distinctive feature is that our estimation and prediction bounds in (\ref{e034})--(\ref{e035}) depend only on the universal regularization parameter $\lambda_0 = c \{(\log p)/n\}^{1/2}$ for the $L_1$-component, and are independent of the regularization parameter $\lambda$ for the concave component. In contrast, the bounds in \cite{ZZ12} generally depend on $\lambda$ alone. The logarithmic factor $\log p$ reflects the general price one needs to pay to search for important variables in high dimensions. In addition, when the signal strength is stronger and the regularization parameter $\lambda$ is chosen suitably,
with the aid of the concave component, we have a stronger variable selection result of sign consistency than using $L_1$-penalty alone, in addition to the oracle inequality. Thanks to the inclusion of the $L_1$-component, another nice feature is that our theory analyzes the sampling properties on the whole parameter space $\mathbb{R}^p$, the full space of all possible models, in contrast to the restriction to the union of lower-dimensional coordinate subspaces such as in \cite{FLv11}.

The bound on the $L_\infty$-estimation loss in Theorem \ref{Thm1} involves $\|(n^{-1} \bX_{1}\t \bX_{1})^{-1}\|_\infty$, which is bounded from above by $s^{1/2} \|(n^{-1} \bX_{1}\t \bX_{1})^{-1}\|_2 \leq s^{1/2} \kappa_0^{-2}$. The former bound is in general tighter than the latter one. To see this, let us consider the special case when all column vectors of the $n\times s$ subdesign matrix $\bX_1$ have equal pairwise correlation $\rho \in [0,1)$. Then the Gram matrix takes the form
$n^{-1} \bX_{1}\t \bX_{1} =(1 - \rho) \bI_s + \rho \bone_s \bone_s\t$. By the Sherman--Morrison--Woodbury formula, we have $(n^{-1}\bX_{1}\t \bX_{1})^{-1}= (1-\rho)^{-1}I_s - \rho(1-\rho)^{-1}\{1+(s-1)\rho\}^{-1}\bone_s \bone_s\t$,
which gives
\[
\|(n^{-1}\bX_{1}\t \bX_{1})^{-1}\|_\infty = (1-\rho)^{-1} [1+ \rho(s-2) \{1+(s-1)\rho\}^{-1}] \leq 2 (1-\rho)^{-1}.
\]
It is interesting to observe that the above matrix $\infty$-norm has a dimension-free upper bound. Thus in this case, the bound on $L_\infty$-estimation loss becomes $O[\{(\log p)/n\}^{1/2}]$.

Due to the presence of the $L_1$-penalty in (\ref{e006}), the resulting global minimizer $\hbbeta$ characterized in Theorem \ref{Thm1} may not have the oracle property in the context of \cite{FL01}. This issue can be resolved using the refitted least-squares estimator on the support $\supp(\hbbeta)$.

\begin{corollary} \label{Cor1}
Assume that all conditions of Theorem \ref{Thm1} hold, and let $\tbbeta$ be the refitted least-squares estimator given by covariates in $\supp(\hbbeta)$, with $\hbbeta$ the estimator in Theorem \ref{Thm1}. Then with probability $1 - O(p^{-c_0})$, $\tbbeta$ equals the oracle estimator, and has the oracle property if the oracle estimator is asymptotic normal.
\end{corollary}

Corollary \ref{Cor1} follows immediately from the second part of Theorem \ref{Thm1}. Additional regularity conditions ensuring the asymptotic normality of the oracle estimator can be found in, for example, Theorem 4 in \cite{FLv11}.

\begin{theorem} \label{Thm2}
Assume that conditions of Theorem \ref{Thm1} hold, with $\veps_1, \ldots, \veps_n$ independent and identically distributed as $\veps_0$. Then the regularized estimator $\hbbeta$ in Theorem \ref{Thm1} satisfies that for any $\tau >0$,
\begin{align}
\label{e052}
E \{n^{-1} \|\bX (\hbbeta - \bbeta_0)\|_2^2\} & = O(\kappa^{-2} \lambda_0^2 s + m_{2, \tau} + \gamma \lambda_0 p^{-c_0}), \\
\nonumber
E (\|\hbbeta - \bbeta_0\|_q^q) & = O[\kappa^{-2 q} \lambda_0^q s + (2 - q) \lambda_0^{-1} m_{2, \tau} + (q - 1) \lambda_0^{-2} m_{4, \tau} \\
\label{e054}
& \quad + \{(2 - q) \gamma + (q - 1) \gamma^2\} p^{-c_0}], \quad q \in [1, 2], \\
\label{e055}
E \{\emph{\mbox{FS}}(\hbbeta)\} & = O\{\kappa^{-4} (\lambda_0/\lambda)^2 s + \lambda^{-2} m_{2, \tau} + (\gamma \lambda_0/\lambda^2 + s) p^{-c_0}\},
\end{align}
where $m_{q, \tau} = E (|\veps_0|^q 1_{\{|\veps_0| > \tau\}})$ denotes
tail moment and $\gamma = \|\bbeta_0\|_1 + s \lambda_0^{-1} p_\lambda(\infty) + \tau^2 \lambda_0^{-1}$. If in addition $\lambda \geq 56 (1- c_1)^{-1} \kappa^{-2} \lambda_0 s^{1/2}$, then we also have $E \{\emph{\mbox{FS}}(\hbbeta)\} = O\{\lambda^{-2} m_{2, \tau} + (\gamma \lambda_0/\lambda^2 + s) p^{-c_0}\}$ and
$E (\|\hbbeta - \bbeta_0\|_\infty) =  O\{\lambda_0 \|(n^{-1} \bX_{1}\t \bX_{1})^{-1}\|_\infty + \lambda_0^{-1} m_{2, \tau} + \gamma p^{-c_0}\}$.
\end{theorem}

Observe that $\lambda_0$ enters all bounds for the oracle risk inequalities, whereas $\lambda$ enters only the risk bound for the variable selection loss. This again reflects the different roles played by the $L_1$-penalty and concave penalty in prediction and variable selection. The estimation and prediction risk bounds in (\ref{e052})--(\ref{e054}) as well as the variable selection risk bound in (\ref{e055}) can have leading orders given in their first terms. To understand this, note that each of these first terms is independent of $\tau$ and $p^{-c_0}$, and the remainders in each upper bound can be sufficiently small, since $\tau$ and $c_0$ can be chosen arbitrarily large. In fact, for bounded error $\veps_i$ with range $[-b, b]$, taking $\tau = b$ makes the tail moments $m_{q,\tau}$ vanish. For Gaussian error $\veps_i \sim N(0, \sigma^2)$, by the Gaussian tail probability bound, we can show that $m_{q,
\tau} = O[\tau^{q - 1} \exp\{-\tau^2/(2 \sigma^2)\}]$ for positive integer $q$. In general, the tail moments can have sufficiently small order by taking a sufficiently large $\tau$ diverging with $n$. All terms involving $p^{-c_0}$ can also be of sufficiently small order by taking a sufficiently large positive constant $c$ in $\lambda_0$; see (\ref{e024}).

Our new oracle risk inequalities complement the common results on the oracle inequalities for losses. The inclusion of the $L_1$-component $\lambda_0 t$ stabilizes prediction and variable selection, and leads to oracle risk bounds. It is, however, unclear whether the concave method alone can enjoy similar risk bounds.

\subsection{Asymptotic properties of computable solutions} \label{Sec3.4}
In \S\ref{Sec3.3} we have shown that the global minimizer for combined $L_1$ and concave regularization can enjoy the appealing asymptotic properties. Such a global minimizer, however, may not be guaranteed to be found by a computational algorithm due to the general nonconvexity of the objective function in (\ref{e006}). Thus a natural question is whether these nice properties can be shared by the computable solution by any algorithm, where a computable solution is typically a local minimizer. Zhang \& Zhang (2012) showed that under regularity conditions, any two sparse local solutions can be close to each other. This result along with the sparsity of the global minimizer in Theorem \ref{Thm1} entails that any sparse computable solution, in the sense of being a local minimizer, can be close to the global minimizer, and thus can enjoy properties similar to the global minimizer. The following theorem establishes these results for sparse computable solutions.

\begin{theorem} \label{Thm3}
Let $\hbbeta$ be a computable local minimizer of (\ref{e006}) that is a global minimizer in each coordinate produced by any algorithm satisfying $\|\hbbeta\|_0 \leq c_2 s$ and $\|n^{-1} \bX\t (\by - \bX \hbbeta)\|_\infty = O(\lambda_0)$, $\lambda \geq c_3 \lambda_0$, and $\min_{\|\bdelta\|_2 = 1,\ \|\bdelta\|_0 \leq c_4 s}n^{-1/2} \|\bX \bdelta\|_2 \geq \kappa_0$ for some positive constants $c_2, c_3, \kappa_0$ and sufficiently large positive constant $c_4$. Then under conditions of Theorem \ref{Thm1}, $\hbbeta$ has the same asymptotic properties as for the global minimizer in theorem \ref{Thm1}.
\end{theorem}

For practical implementation of method in (\ref{e006}), we employ the path-following coordinate optimization algorithm (Fan \& Lv, 2011; Mazumder et al., 2011) and choose the initial estimate as the lasso estimator $\hbbeta_{\text{lasso}}$ with the regularization parameter tuned to minimize the cross-validated prediction error. An analysis of the convergence properties of such an algorithm was presented by \cite{LL13}. The use of the lasso estimator as the initial value has also been exploited in, for example, Zhang \& Zhang (2012). With the coordinate optimization algorithm, one can obtain a path of sparse computable solutions that are global minimizers in each coordinate. Theorem \ref{Thm3} suggests that a sufficiently sparse computable solution with small correlation between the residual vector and all covariates can enjoy desirable properties.

\section{A simulation study} \label{Sec4}

We simulated 100 data sets from the linear regression model (\ref{e001}) with $\bveps \sim N(\bzero,
\sigma^2 I_n)$ and $\sigma = $ 0$\cdot$25. For each simulated data set, the rows of $\bX$ were sampled as independent and identically distributed copies from $N(\bzero, \Sigma_0)$ with $\Sigma_0 = ($0$\cdot$5$^{|i-j|})$. We considered $(n, p) = (80, 1000)$ and $(160, 4000)$, and set $\bbeta$ as $\bbeta_0 = (1, -$0$\cdot$5$, $ 0$\cdot$7$, -$1$\cdot$2$, -$0$\cdot$9$, $ 0$\cdot$3$, $ 0$\cdot$55$, 0, \ldots, 0)\t$. For each data set, we employed the lasso, combined $L_1$ and the smoothly clipped absolute deviation \citep{FL01}, combined $L_1$ and hard-thresholding, and combined $L_1$ and the smooth integration of counting and absolute deviation penalties to produce a sparse estimate. The minimax concave penalty in \cite{Zhang10} performed very similarly to the smoothly clipped absolute deviation penalty, so we omit its results to save space. The tuning parameters were selected using BIC.

\begin{table}
\def~{\hphantom{0}}
\tbl{Means and standard errors (in parentheses) of different performance measures}{%
\begin{tabular}{lccccc}
 \\
& Lasso & $L_1$+SCAD & $L_1$+Hard & $L_1$+SICA & Oracle \\[5pt]
$n = 80$ &  &  & &  & \\
PE ($\times 10^{-2}$) & 45$\cdot$0 (1$\cdot$7) & 8$\cdot$1 (0$\cdot$2) & 7$\cdot$0 (0$\cdot$1) & 7$\cdot$1 (0$\cdot$1) & 6$\cdot$9 (0$\cdot$0) \\
$L_2$-loss ($\times 10^{-2}$) & 86$\cdot$9 (1$\cdot$9) & 16$\cdot$8 (1$\cdot$0) & 11$\cdot$3 (0$\cdot$4) & 11$\cdot$3 (0$\cdot$5) & 9$\cdot$7 (0$\cdot$3) \\
$L_1$-loss ($\times 10^{-1}$) & 27$\cdot$6 (0$\cdot$6) & 3$\cdot$6 (0$\cdot$2) & 2$\cdot$5 (0$\cdot$1) & 2$\cdot$5 (0$\cdot$1) & 2$\cdot$1 (0$\cdot$1) \\
$L_\infty$-loss ($\times 10^{-2}$) & 48$\cdot$2 (1$\cdot$2) & 12$\cdot$1 (0$\cdot$8) & 7$\cdot$5 (0$\cdot$3) & 7$\cdot$5 (0$\cdot$3) & 6$\cdot$6 (0$\cdot$2) \\
FP & 26$\cdot$1 (0$\cdot$5) & 0$\cdot$2 (0$\cdot$0) & 0 (0) & 0 (0) & 0 (0) \\
FN & 1$\cdot$0 (0$\cdot$1) & 0$\cdot$1 (0$\cdot$0) & 0$\cdot$0 (0$\cdot$0) & 0$\cdot$0 (0$\cdot$0) & 0 (0)\\[5pt]
$n = 160$ &  &  & &  & \\
PE ($\times 10^{-2}$) & 16$\cdot$9 (0$\cdot$5) & 6$\cdot$7 (0$\cdot$0) & 7$\cdot$0 (0$\cdot$1) & 7$\cdot$0 (0$\cdot$1) & 6$\cdot$6 (0$\cdot$0) \\
$L_2$-loss ($\times 10^{-2}$) & 45$\cdot$3 (1$\cdot$0) & 7$\cdot$7 (0$\cdot$3) & 9$\cdot$2 (0$\cdot$4) & 9$\cdot$2 (0$\cdot$4) & 6$\cdot$6 (0$\cdot$2) \\
$L_1$-loss ($\times 10^{-1}$) & 16$\cdot$2 (0$\cdot$3) & 1$\cdot$7 (0$\cdot$1) & 2$\cdot$1 (0$\cdot$1) & 2$\cdot$1 (0$\cdot$1) & 1$\cdot$4 (0$\cdot$0) \\
$L_\infty$-loss ($\times 10^{-2}$) & 24$\cdot$9 (0$\cdot$6) & 5$\cdot$3 (0$\cdot$2) & 6$\cdot$0 (0$\cdot$2) & 5$\cdot$9 (0$\cdot$2) & 4$\cdot$4 (0$\cdot$1) \\
FP & 52$\cdot$8 (1$\cdot$1) & 0$\cdot$1 (0$\cdot$0) & 0$\cdot$7 (0$\cdot$1) & 0$\cdot$7 (0$\cdot$1) & 0 (0) \\
FN & 0 (0) & 0 (0) & 0 (0) & 0 (0) & 0 (0)
\end{tabular}}
\label{tab1}
\begin{tabnote}
$L_1$+SCAD, combined $L_1$ and smoothly clipped absolute deviation; $L_1$+Hard, combined $L_1$ and hard-thresholding; $L_1$+SICA, combined $L_1$ and smooth integration of counting and absolute deviation; PE, prediction error; FP, number of false positives; FN, number of false negatives.
\end{tabnote}
\end{table}

We considered six performance measures for the estimate $\hbbeta$: the prediction error, $L_2$-loss, $L_1$-loss, $L_\infty$-loss, the number of false positives, and the number of false negatives. The prediction error is defined as $E (Y - \bx\t \hbbeta)^2$, with $(\bx\t, Y)$ an independent observation, which was calculated based on an independent test sample of size 10,000. The $L_q$-loss for estimation is $\|\hbbeta - \bbeta_0\|_q$. A false positive means a selected covariate outside the true sparse model $\supp(\bbeta_0)$, and a false negative means a missed covariate in $\supp(\bbeta_0)$.

Table \ref{tab1} lists the results under different performance measures. The combined $L_1$ and smoothly clipped absolute deviation, combined $L_1$ and hard-thresholding, and combined $L_1$ and smooth integration of counting and absolute deviation all performed similarly to the oracle procedure, outperforming the lasso. When the sample size increases, the performance of all methods tends to improve. Although theoretically the oracle inequalities for the $L_1$-penalty and combined $L_1$ and concave penalty can have the same convergence rates, the constants in these oracle inequalities matter in finite samples. This explains the differences in prediction errors and other performance measures in Table \ref{tab1} for various methods.

We also compared our method with the concave penalty alone. Simulation studies suggest that they have similar performance, except that our method is more stable. To illustrate this, we compared the smoothly clipped absolute deviation with combined $L_1$ and the smoothly clipped absolute deviation. Boxplots of different performance measures by the two methods showed that the latter reduces the outliers and variability, and thus stabilizes the estimate. This result reveals that the same advantage as advocated in \cite{LW07} remains true in high dimensions, with more general concave penalties.

\section{Real data analysis} \label{Sec5}

We applied our method to the lung cancer data originally studied in Gordon et al. (2002) and analyzed in Fan \& Fan (2008). This consists of 181 tissue samples, with 31 from the malignant pleural mesothelioma of the lung, and 150 from the adenocarcinoma of the lung. Each sample tissue is described by 12533 genes.

To better evaluate the suggested method, we randomly split the 181 samples into a training set and a test set such that the training set consists of 16 samples from the malignant pleural mesothelioma class and 75 samples from the adenocarcinoma class. Correspondingly, the test set has 15 samples from the malignant pleural mesothelioma class and 75 samples from the adenocarcinoma class. For each split, we employed the same methods as in \S\ref{Sec4} to fit the logistic regression model to the training data, and then calculated the classification error using the test data. The tuning parameters were selected using the cross-validation. We repeated the random splitting 50 times, and the means and standard errors of classification errors were 2$\cdot$960 (0$\cdot$254) for the lasso, 3$\cdot$080 (0$\cdot$262) for combined $L_1$ and the smoothly clipped absolute deviation, 2$\cdot$960 (0$\cdot$246) for combined $L_1$ and hard-thresholding, and 2$\cdot$980 (0$\cdot$228) for combined $L_1$ and the smooth integration of counting and absolute deviation. We also calculated the median number of variables chosen by each method: 19 for the first one, 11 for the second one, 11 for the third one, and 12 for the fourth one; the mean model sizes are almost the same as the medians. For each method, we computed the percentage of times each gene was selected, and list the most frequently chosen $m$ genes in the Supplementary Material, with $m$ equal to the median model size by the method. The sets of genes selected by the combined $L_1$ and concave penalties are subsets of those selected by the lasso.

\section{Discussion} \label{Sec6}
Our theoretical analysis shows that the regularized estimate, as the global optimum, given by combined $L_1$ and concave regularization enjoys the same asymptotic properties as the lasso estimator, but with improved sparsity and false sign rate, in ultra-high dimensional linear regression model. These results may be extended to more general model settings and other convex penalties, such as the $L_2$-penalty. To quantify the stability of variable selection, one can use, for example, the bootstrap method \citep{Efron79} to estimate the selection probabilities, significance, and estimation uncertainty of selected variables by the regularization method in practice.

\section*{Acknowledgement}
The authors sincerely thank the editor, an associate editor, and two referees for comments that significantly improved the paper. This work was supported by the U.S. National Science Foundation and the University of Southern California.

\section*{Supplementary material}
\label{SM}
Supplementary material available at \Bka\ online includes the proofs of Proposition \ref{Prop1} and Theorem \ref{Thm3}, and further details for \S\ref{Sec5}.

\appendix

\appendixone
\section*{Appendix 1}

\subsection*{Proof of Theorem \ref{Thm1}} \label{SecA.2}
Let $\bdelta = \hbbeta - \bbeta_0$ denote the estimation error with $\hbbeta$ the global minimizer of (\ref{e006}). By Condition \ref{cond2}, we see from Proposition \ref{Prop1} that each $\hbeta_j$ is either 0 or of magnitude larger than $(1 - c_1) \lambda$.
It follows from the global optimality of $\hbbeta$ that
\begin{equation} \label{e003}
(2n)^{-1} \|\bveps - \bX (\hbbeta - \bbeta_0)\|_2^2 + \lambda_0 \|\hbbeta\|_1 + \|p_\lambda(\hbbeta)\|_1 \leq (2n)^{-1} \|\bveps\|_2^2 + \lambda_0 \|\bbeta_0\|_1 + \|p_\lambda(\bbeta_0)\|_1.
\end{equation}
With some simple algebra, (\ref{e003}) becomes
\begin{equation} \label{e004}
(2n)^{-1} \|\bX \bdelta\|_2^2 - n^{-1} \bveps\t \bX \bdelta + \lambda_0 \|\bbeta_0 + \bdelta\|_1 + \|p_\lambda(\bbeta_0 + \bdelta)\|_1 \leq \lambda_0 \|\bbeta_0\|_1 + \|p_\lambda(\bbeta_0)\|_1.
\end{equation}
For notational simplicity, we let $\widetilde{\ba}_1$ and $\widetilde{\ba}_2$ denote the subvectors of a $p$-vector $\ba$ consisting of its first $s$ components and remaining $p - s$ components, respectively. Since $\widetilde{\bbeta}_{0,2} = \bzero$, we have $\widetilde{\bbeta}_{0,2} + \widetilde{\bdelta}_2=\widetilde{\bdelta}_2$. Thus we can rewrite (\ref{e004}) as
\begin{equation} \label{e005}
(2n)^{-1} \|\bX \bdelta\|_2^2 - n^{-1} \bveps\t \bX \bdelta + \lambda_0 \|\widetilde{\bdelta}_{2}\|_1 \leq \lambda_0 \|\widetilde{\bbeta}_{0, 1}\|_1 - \lambda_0 \|\widetilde{\bbeta}_{0, 1} + \widetilde{\bdelta}_{1}\|_1 + \|p_\lambda(\bbeta_0)\|_1 - \|p_\lambda(\bbeta_0 + \bdelta)\|_1.
\end{equation}
The reverse triangle inequality $|\lambda_0 \|\widetilde{\bbeta}_{0, 1}\|_1 - \lambda_0 \|\widetilde{\bbeta}_{0,1} + \widetilde{\bdelta}_{1}\|_1| \leq \lambda_0 \|\widetilde{\bdelta}_{1}\|_1$ along with (\ref{e005}) yields
\begin{equation} \label{e014}
(2n)^{-1} \|\bX \bdelta\|_2^2 - n^{-1} \bveps\t \bX \bdelta + \lambda_0 \|\widetilde{\bdelta}_{2}\|_1 \leq \lambda_0 \|\widetilde{\bdelta}_{1}\|_1 + \|p_\lambda(\bbeta_0)\|_1 - \|p_\lambda(\bbeta_0 + \bdelta)\|_1,
\end{equation}
which is key to establishing bounds on prediction and variable selection losses.

To analyze the behavior of $\bdelta$, we need to use the concentration property of $n^{-1} \bX\t \bveps$ around its mean zero, as given in the deviation probability bound (\ref{e024}). Condition on the event $\mathscr{E} = \{\|n^{-1} \bX\t \bveps\|_\infty \leq \lambda_0/2\}$. On this event, we have
\[
- n^{-1} \bveps\t \bX \bdelta + \lambda_0 \|\widetilde{\bdelta}_{2}\|_1 - \lambda_0 \|\widetilde{\bdelta}_{1}\|_1 \geq -(\lambda_0/2) \|\bdelta\|_1 + \lambda_0 \|\widetilde{\bdelta}_{2}\|_1 - \lambda_0 \|\widetilde{\bdelta}_{1}\|_1  = (\lambda_0/2) \|\widetilde{\bdelta}_{2}\|_1 - (3 \lambda_0/2) \|\widetilde{\bdelta}_{1}\|_1.
\]
This inequality together with (\ref{e014}) gives
\begin{equation} \label{e011}
(2n)^{-1} \|\bX \bdelta\|_2^2 + (\lambda_0/2) \|\widetilde{\bdelta}_{2}\|_1 \leq (3 \lambda_0/2) \|\widetilde{\bdelta}_{1}\|_1 + \|p_\lambda(\bbeta_0)\|_1 - \|p_\lambda(\bbeta_0 + \bdelta)\|_1.
\end{equation}
In order to proceed, we need to construct an upper bound for $\|p_\lambda(\bbeta_0)\|_1 - \|p_\lambda(\bbeta_0 + \bdelta)\|_1$. We claim that such an upper bound is $(4n)^{-1} \|\bX \bdelta\|_2^2 + 4^{-1} \lambda_0 \|\bdelta\|_1$. To prove this, we consider two cases.

\textit{Case 1}: $\|\hbbeta\|_0 \geq s$. Then by Condition \ref{cond2}, we have $|\beta_{0, j}| > (1 - c_1) \lambda$ ($j = 1,\ldots, s$) and $p_\lambda'\{(1 - c_1) \lambda\} \leq \lambda_0/4$. For each $j = 1,\ldots, s$, if $\hbeta_j \neq 0$, we must have $|\hbeta_j| > (1 - c_1) \lambda$ and thus by the mean-value theorem,
$ |p_\lambda(|\beta_{0, j}|) - p_\lambda(|\hbeta_j|)| = p_\lambda'(t) |(|\hbeta_j| - |\beta_{0, j}|)| \leq p_\lambda'(t) |\delta_j|$,
where $t$ is between $|\beta_{0, j}|$ and $|\hbeta_j|$, and $\delta_j$ is the $j$th component of $\bdelta$. Clearly $t > (1 - c_1) \lambda$, which along with the concavity of $p_\lambda(t)$ leads to $p_\lambda'(t) \leq p_\lambda'\{(1 - c_1) \lambda\} \leq \lambda_0/4$. This shows that $|p_\lambda(|\beta_{0, j}|) - p_\lambda(|\hbeta_j|)| \leq 4^{-1} \lambda_0 |\delta_j|$ for each $j = 1, \ldots, s$ with $\hbeta_j \neq 0$. We now consider $j = 1,\ldots, s$ with $\hbeta_j = 0$. Since $\|\hbbeta\|_0 \geq s$, there exists some $j' > s$ such that $\hbeta_{j'} \neq 0$ and $j'$'s are distinct for different $j$'s. Similarly as above, we have for some $t_1$ between $(1 - c_1) \lambda$ and $|\beta_{0, j}|$ and some $t_2$ between $(1 - c_1) \lambda$ and $|\hbeta_{j'}|$,
\begin{align*}
|p_\lambda(|\beta_{0, j}|) - p_\lambda(|\hbeta_{j'}|)| & \leq |p_\lambda(|\beta_{0, j}|) - p_\lambda\{(1 - c_1) \lambda\}| + |p_\lambda(|\hbeta_{j'}|) - p_\lambda\{(1 - c_1) \lambda\}|  \\
& = p_\lambda'(t_1) \{|\beta_{0, j}| - (1 - c_1) \lambda\} + p_\lambda'(t_2) \{|\hbeta_{j'}| - (1 - c_1) \lambda\} \\
&\leq p_\lambda'(t_1) |\delta_j| + p_\lambda'(t_2) |\delta_{j'}| \leq (\lambda_0/4) (|\delta_j| + |\delta_{j'}|),
\end{align*}
since $\hbeta_j = 0$ and $\beta_{0, j'} = 0$. Combining these two sets of inequalities yields the desired upper bound
$ \|p_\lambda(\bbeta_0)\|_1 - \|p_\lambda(\bbeta_0 + \bdelta)\|_1 \leq  (\lambda_0/4) \|\bdelta\|_1 \leq (4n)^{-1} \|\bX \bdelta\|_2^2 + \lambda_0 \|\bdelta\|_1/4$.

\textit{Case 2}: $\|\hbbeta\|_0 = s - k$ for some $k \geq 1$. Then we have $\|\bdelta\|_0 \leq \|\hbbeta\|_0 + \|\bbeta_0\|_0 \leq s - k + s < 2 s$ and $\|\bdelta\|_2 \geq k^{1/2} \min_{j = 1, \ldots, s} |\beta_{0, j}|$, since there are at least $k$ such $j$'s with $j = 1, \ldots, s$ and $\hbeta_j = 0$. Thus it follows from the first part of Condition \ref{cond1} and $\min_{j = 1, \ldots, s} |\beta_{0, j}| > 2 \kappa_0^{-1} p_\lambda^{1/2}(\infty)$ in Condition \ref{cond2} that
\[ (4n)^{-1} \|\bX \bdelta\|_2^2 \geq 4^{-1} \kappa_0^2 \|\bdelta\|_2^2 \geq 4^{-1} \kappa_0^2 (k^{1/2} \min_{j = 1, \ldots, s} |\beta_{0, j}|)^2 \geq k p_\lambda(\infty).  \]
Since $p_\lambda(|\beta_{0, j}|) \leq p_\lambda(\infty)$ and there are $s - k$ nonzero $\hbeta_j$'s, applying the same arguments as in Case 1 gives our desired upper bound
$ \|p_\lambda(\bbeta_0)\|_1 - \|p_\lambda(\bbeta_0 + \bdelta)\|_1 \leq  k p_\lambda(\infty) + (\lambda_0/4) \|\bdelta\|_1 \leq (4n)^{-1} \|\bX \bdelta\|_2^2 + (\lambda_0/4) \|\bdelta\|_1$.

Combining Cases 1 and 2 above along with (\ref{e011}) and $\|\bdelta\|_1 = \|\widetilde{\bdelta}_1\|_1 + \|\widetilde{\bdelta}_2\|_1$ yields
\begin{equation} \label{e015}
n^{-1} \|\bX \bdelta\|_2^2 + \lambda_0 \|\widetilde{\bdelta}_{2}\|_1 \leq 7 \lambda_0 \|\widetilde{\bdelta}_{1}\|_1,
\end{equation}
which entails a basic constraint
\begin{equation} \label{e012}
\|\widetilde{\bdelta}_{2}\|_1 \leq 7 \|\widetilde{\bdelta}_{1}\|_1.
\end{equation}
With the aid of (\ref{e012}), we will first establish a useful bound on $\|\widetilde{\bdelta}_{2}\|_2$. In view of (\ref{e012}), the restricted eigenvalue assumption in the second part of Condition \ref{cond1} and (\ref{e015}), as well as the Cauchy--Schwartz inequality, lead to
\begin{equation} \label{e027}
4^{-1} \kappa^2(s, 7) (\|\widetilde{\bdelta}_{1}\|_2^2 \vee \|\widetilde{\bdelta}'_2\|_2^2) \leq (4n)^{-1} \|\bX \bdelta\|_2^2 \leq (7/4) \lambda_0 \|\widetilde{\bdelta}_{1}\|_1 \leq (7/4) \lambda_0 s^{1/2} \|\widetilde{\bdelta}_{1}\|_2.
\end{equation}
Solving this inequality gives
\begin{equation} \label{e016}
\|\widetilde{\bdelta}_{1}\|_2 \leq 7 \lambda_0 s^{1/2}/\kappa^2(s, 7), \quad \|\widetilde{\bdelta}_{1}\|_1\leq s^{1/2}\|\widetilde{\bdelta}_{1}\|_2 \leq 7\lambda_0s/\kappa^2(s, 7).
\end{equation}
Since the $k$th largest absolute component of $\widetilde{\bdelta}_{2}$ is bounded from above by $\|\widetilde{\bdelta}_{2}\|_1/k$, we have
$ \|\widetilde{\bdelta}_3\|_2^2 \leq \sum_{k = s + 1}^{p-s} \|\widetilde{\bdelta}_{2}\|_1^2/k^2 \leq s^{-1} \|\widetilde{\bdelta}_{2}\|_1^2$,
where $\widetilde{\bdelta}_3$ is a subvector of $\widetilde{\bdelta}_2$ consisting of components excluding those with the $s$ largest magnitude. This inequality, (\ref{e012}), and the Cauchy--Schwartz inequality entail that
\[ \|\widetilde{\bdelta}_3\|_2 \leq s^{-1/2} \|\widetilde{\bdelta}_{2}\|_1 \leq 7 s^{-1/2} \|\widetilde{\bdelta}_{1}\|_1 \leq 7 \|\widetilde{\bdelta}_{1}\|_2, \]
and thus $\|\widetilde{\bdelta}_2\|_2 \leq 7 \|\widetilde{\bdelta}_{1}\|_2 + \|\widetilde{\bdelta}'_2\|_2$. By (\ref{e027}), we have $\|\widetilde{\bdelta}'_2\|_2 \leq 7^{1/2} \lambda_0^{1/2} s^{1/4} \|\widetilde{\bdelta}_{1}\|_2^{1/2}/\kappa(s, 7)$. Combining these two inequalities with (\ref{e016}) gives
\begin{equation} \label{e025}
\|\widetilde{\bdelta}_2\|_2 \leq 7 \|\widetilde{\bdelta}_{1}\|_2 + 7^{1/2} \lambda_0^{1/2} s^{1/4} \|\widetilde{\bdelta}_{1}\|_2^{1/2}/\kappa(s, 7) \leq 56 \lambda_0 s^{1/2}/\kappa^2(s, 7).
\end{equation}
This bound enables us to conduct more delicate analysis on $\bdelta$.

We proceed to prove the first part of Theorem \ref{Thm1}. The inequality (\ref{e034}) on the prediction loss can be obtained by inserting (\ref{e016}) into (\ref{e015}):
\begin{equation}\label{e109}
n^{-1/2}\|\bX\bdelta\|_2 \leq 7 \lambda_0s^{1/2}/\kappa(s,7).
\end{equation}
Combining (\ref{e016}) with (\ref{e025})  yields the following bound on the $L_2$-estimation loss,
\begin{equation}\label{e113}
\|\bdelta\|_2 \leq \|\widetilde{\bdelta}_1\|_2 + \|\widetilde{\bdelta}_2\|_2 \leq  63\lambda_0s^{1/2}/\kappa^2(s,7).
\end{equation}
For each $1 \leq q < 2$, an application of H\"{o}lder's inequality gives
\begin{equation} \label{e112}
\|\bdelta\|_q  \leq \{s^{(2 - q)/2} \|\widetilde{\bdelta}_1\|_2^q\}^{1/q} = s^{(2 - q)/(2 q)} \|\bdelta\|_2 \leq 63 \lambda_0s^{1/q}/\kappa^2(s,7).
\end{equation}
Now we bound the number of falsely discovered signs $\FS(\hbbeta)$. If $\sgn(\hbeta_j)\neq \sgn(\beta_{0,j})$, then by Proposition \ref{Prop1} and Condition 2, $|\delta_j|=|\hbeta_j-\beta_{0,j}|\geq (1-c_1)\lambda$. Thus, it follows that $\|\bdelta\|_2 \geq \{\FS(\hbbeta)\}^{1/2}(1-c_1)\lambda$. This together with (\ref{e113}) entails that
\begin{equation}\label{e007}
\FS(\hbbeta) \leq \{63/(1-c_1)\}^{2}(\lambda_0/\lambda)^2s/\kappa^4(s,7).
\end{equation}
We finally note that all the above bounds for $\hbbeta$ are conditional on the event $\mathscr{E}$, and thus hold simultaneously with probability $1 - O(p^{-c_0})$, which concludes the proof for the first part of Theorem \ref{Thm1}.

It remains to prove the second part of Theorem \ref{Thm1}.
Since $\lambda \geq 56 (1- c_1)^{-1} \lambda_0 s^{1/2}/\kappa^2(s, 7)$, we have by Condition \ref{cond2} that
$\min_{j = 1, \ldots, s} |\beta_{0, j}| >  56 \lambda_0 s^{1/2}/\kappa^2(s, 7)$. This inequality together with (\ref{e016}) entails that for each $ j =1,\ldots, s$,
\begin{equation} \label{e037}
\sgn(\hbeta_{j}) = \sgn(\beta_{0, j}),
\end{equation}
by a simple contradiction argument. In view of (\ref{e025}) and the hard-thresholding feature of $\hbbeta = (\hbbeta_{0,1}\t, \hbbeta_{0,2}\t)\t$ with $\hbbeta_{0,1} = (\hbeta_1, \ldots, \hbeta_s)\t$, a similar contradiction argument shows that $\hbbeta_{0,2} = \bzero$. Combining this result with (\ref{e037}) leads to $\sgn(\hbbeta) = \sgn(\bbeta_0)$. With this strong result on sign consistency of $\hbbeta$, we can derive tight bounds on the $L_\infty$-loss. By Theorem 1 of Lv \& Fan (2009), $\hbbeta_{0,1}$ solves the following equation for $\bgamma \in \mathbb{R}^s$
\begin{equation} \label{e036}
\bgamma = \widetilde{\bbeta}_{0,1} - (n^{-1} \bX_{1}\t \bX_{1})^{-1} \bb,
\end{equation}
where $\bX_{1}$ is an $n \times s$ submatrix of $\bX$ corresponding to $s$ nonzero $\beta_{0, j}$'s and $\bb = \{\lambda_0 \bone_s + p_\lambda'(|\bgamma|)\} \circ \sgn(\widetilde{\bbeta}_{0, 1}) - n^{-1} \bX_{1}\t \bveps$, with the derivative taken componentwise and $\circ$ the Hadamard, componentwise, product. It follows from the concavity and monotonicity of $p_\lambda(t)$ and Condition \ref{cond2} that for any $t > (1 - c_1) \lambda$, we have $0 \leq p_\lambda'(t) \leq p_\lambda'((1 - c_1) \lambda) \leq \lambda_0/4$. In view of (\ref{e037}) and the hard-thresholding feature of $\hbbeta$, each component of $\hbbeta_{0,1}$ has magnitude larger than $(1 - c_1) \lambda$. Since $\|n^{-1} \bX_{1}\t \bveps\|_\infty \leq \|n^{-1} \bX\t \bveps\|_\infty \leq \lambda_0/2$ on the event $\mathscr{E}$, combining these results leads to
\begin{equation} \label{e038}
\sgn(\bb) = \sgn(\widetilde{\bbeta}_{0,1}), \quad \lambda_0/2 \leq \|\bb\|_\infty \leq 7 \lambda_0/4.
\end{equation}
Clearly $\widetilde{\bdelta}_{2} = \hbbeta_{0,2} = \bzero$. Thus it follows from (\ref{e036}), (\ref{e038}), and the first part of Condition \ref{cond1} that
\begin{align}
\label{e111}
\|\bdelta\|_\infty & \leq  \|(n^{-1} \bX_{1}\t \bX_{1})^{-1}\|_\infty \|\bb\|_\infty \leq (7/4) \lambda_0 \|(n^{-1} \bX_{1}\t \bX_{1})^{-1}\|_\infty,
\end{align}
which concludes the proof for the second part of Theorem \ref{Thm1}.

\subsection*{Proof of Theorem \ref{Thm2}} \label{SecA.3}
Let $\hbbeta$ be the global minimizer of (\ref{e006}) given in Theorem \ref{Thm1}, with $\bdelta = \hbbeta - \bbeta_0$ denoting the estimation error. To calculate the risk of the regularized estimator $\hbbeta$ for different losses, we need to analyze its tail behavior on the event $\mathscr{E}^c = \{\|n^{-1} \bX\t \bveps\|_\infty > \lambda_0/2\}$.
We work directly with inequality (\ref{e003}). It follows easily from (\ref{e003}) that
\begin{equation} \label{e040}
(2n)^{-1} \|\bX \bdelta - \bveps\|_2^2 + \lambda_0 \|\bdelta\|_1 + \|p_\lambda(\hbbeta)\|_1 \leq (2n)^{-1} \|\bveps\|_2^2 + 2 \lambda_0 \|\bbeta_0\|_1 + \|p_\lambda(\bbeta_0)\|_1.
\end{equation}
We need to bound the term $E\{(2n)^{-1} \|\bveps\|_2^2 1_{\mathscr{E}^c}\}$ from above, where $\bveps = (\veps_1, \ldots, \veps_n)\t$. Consider the cases of bounded or unbounded error.

\textit{Case 1}: Bounded error with range $[-b, b]$. Then in view of the deviation probability bound (\ref{e024}), we have
\begin{equation} \label{e041}
E\{(2n)^{-1} \|\bveps\|_2^2 1_{\mathscr{E}^c}\} \leq (b^2/2) {\rm pr}(\mathscr{E}^c) = O(p^{-c_0}).
\end{equation}

\textit{Case 2}: Unbounded error. Then it follows from (\ref{e024}) that for each $i = 1, \ldots, n$ and any $\tau > 0$,
\begin{align*}
E\{\veps_i^2 1_{\mathscr{E}^c}\} & \leq E\{\veps_i^2 1_{\{|\veps_i| \leq \tau\} \cap \mathscr{E}^c}\} + E\{\veps_i^2 1_{\{|\veps_i| > \tau\}}\} \leq \tau^2 {\rm pr}(\mathscr{E}^c) + E\{\veps_0^2 1_{\{|\veps_0| > \tau\}}\} \\
&= O(\tau^2 p^{-c_0}) + E\{\veps_0^2 1_{\{|\veps_0| > \tau\}}\}.
\end{align*}
Thus we have
\begin{equation} \label{e042}
E \{(2n)^{-1} \|\bveps\|_2^2 1_{\mathscr{E}^c}\} = (2n)^{-1} \sum_{i = 1}^n E\{ \veps_i^2 1_{\mathscr{E}^c}\}
\leq 2^{-1} E\{\veps_0^2 1_{\{|\veps_0| > \tau\}}\} + O(\tau^2 p^{-c_0}).
\end{equation}
Clearly, the bound (\ref{e041}) is a special case of the general bound (\ref{e042}), with $\tau = b$.

We first consider the risks under the $L_1$-loss and prediction loss. Note that $\|p_\lambda(\bbeta_0)\|_1 \leq s p_\lambda(\infty)$. By (\ref{e040}), (\ref{e042}), and (\ref{e024}), we have
\begin{align}
\nonumber
E \{\|\bdelta\|_1 1_{\mathscr{E}^c}\} & \leq \lambda_0^{-1} E\{(2n)^{-1} \|\bveps\|_2^2 1_{\mathscr{E}^c}\} + O[\{2 \|\bbeta_0\|_1 + s \lambda_0^{-1} p_\lambda(\infty)\} p^{-c_0}] \\
\label{e044}
& \leq (2 \lambda_0)^{-1} E\{\veps_0^2 1_{\{|\veps_0| > \tau\}}\} + O(\gamma p^{-c_0}),
\end{align}
where $\gamma = \|\bbeta_0\|_1 + s \lambda_0^{-1} p_\lambda(\infty) + \tau^2 \lambda_0^{-1}$.
This inequality along with (\ref{e112}) on the event $\mathscr{E}$ yields for any $\tau > 0$,
\[
E \|\bdelta\|_1 \leq 63\lambda_0s/\kappa^2(s,7)  + (2 \lambda_0)^{-1} E\{\veps_0^2 1_{\{|\veps_0| > \tau\}}\} + O(\gamma p^{-c_0}).
\]
Note that $(2n)^{-1} \|\bX \bdelta - \bveps\|_2^2 \geq (4n)^{-1} \|\bX \bdelta\|_2^2 - (2n)^{-1} \|\bveps\|_2^2$. Thus in view of (\ref{e040}), a similar argument as for (\ref{e044}) applies to show that $E \{n^{-1} \|\bX \bdelta\|_2^2 1_{\mathscr{E}^c}\} \leq 4 E\{\veps_0^2 1_{\{|\veps_0| > \tau\}}\} + O(\gamma \lambda_0 p^{-c_0})$. Combining this inequality with (\ref{e109}) on the event $\mathscr{E}$ gives
\[ E \{n^{-1} \|\bX \bdelta\|_2^2\} \leq 49\lambda_0^2s/\kappa^2(s,7) + 4 E\{\veps_0^2 1_{\{|\veps_0| > \tau\}}\} + O(\gamma \lambda_0 p^{-c_0}). \]

We now consider the risk under the variable selection loss. To this end, we need to bound $\|\hbbeta\|_0$ on the event $\mathscr{E}^c$. Since $\hbbeta$ always has the hard-thresholding property ensured by Proposition \ref{Prop1}, it follows from the monotonicity of $p_\lambda(t)$ and Condition \ref{cond2} that
$
\|p_\lambda(\hbbeta)\|_1  \geq \|\hbbeta\|_0 p_\lambda((1 - c_1) \lambda) \geq \|\hbbeta\|_0 p_{\text{H}, \lambda}((1 - c_1) \lambda) = \|\hbbeta\|_0 2^{-1} (1 - c_1^2) \lambda^2$.
This inequality along with (\ref{e040}) shows
\begin{equation} \label{e045}
\|\hbbeta\|_0 \leq 2 (1 - c_1^2)^{-1} \lambda^{-2} \{(2n)^{-1} \|\bveps\|_2^2 + 2 \lambda_0 \|\bbeta_0\|_1 + \|p_\lambda(\bbeta_0)\|_1\}.
\end{equation}
Clearly, $\mbox{FS}(\hbbeta) \leq \|\hbbeta\|_0 + s$. Thus by (\ref{e045}), applying a similar argument as for (\ref{e044}) gives
\begin{equation} \label{e072}
E \{\mbox{FS}(\hbbeta) 1_{\mathscr{E}^c}\} \leq (1 - c_1^2)^{-1} \lambda^{-2} E\{\veps_0^2 1_{\{|\veps_0| > \tau\}}\} + O\{(\gamma \lambda_0/\lambda^2 + s) p^{-c_0}\}.
\end{equation}
It follows from this bound and inequality (\ref{e007}) on the event $\mathscr{E}$ that
\[
E \{\mbox{FS}(\hbbeta)\} \leq 63^2(1-c_1)^{-2}(\lambda_0/\lambda)^2s/\kappa^4(s,7)+ (1 - c_1^2)^{-1} \lambda^{-2} E\{\veps_0^2 1_{\{|\veps_0| > \tau\}}\} + O\{(\gamma \lambda_0/\lambda^2 + s) p^{-c_0}\}.
\]

We finally consider the risks under the $L_q$-loss with $q \in (1, 2]$. By (\ref{e040}) and the norm inequality $\|\bdelta\|_2 \leq \|\bdelta\|_1$, we have
\begin{align*}
\|\bdelta\|_2^2 & \leq \lambda_0^{-2} \{(2n)^{-1} \|\bveps\|_2^2 + 2 \lambda_0 \|\bbeta_0\|_1 + \|p_\lambda(\bbeta_0)\|_1\}^2 \leq 3 \lambda_0^{-2} \{(2n)^{-2} \|\bveps\|_2^4 + 4 \lambda_0^2 \|\bbeta_0\|_1^2 + \|p_\lambda(\bbeta_0)\|_1^2\} \\
& \leq 3 \lambda_0^{-2} \big\{(4 n)^{-1} \sum\nolimits_{i = 1}^n \veps_i^4 + 4 \lambda_0^2 \|\bbeta_0\|_1^2 + s^2 p_\lambda^2(\infty)\big\}.
\end{align*}
With this inequality and (\ref{e024}), a similar argument as for (\ref{e042}) applies to show that for any $\tau > 0$,
\begin{align}
\nonumber
E \{\|\bdelta\|_2^2 1_{\mathscr{E}^c}\} & \leq 3 \lambda_0^{-2} \big[(4 n)^{-1} \sum\nolimits_{i = 1}^n E (\veps_i^4 1_{\mathscr{E}^c}) + \{4 \lambda_0^2 \|\bbeta_0\|_1^2 + s^2 p_\lambda^2(\infty)\} {\rm pr}(\mathscr{E}^c) \big] \\
\label{e046}
& \leq (3/4) \lambda_0^{-2} E(\veps_0^4 1_{\{|\veps_0| > \tau\}}) + O(\gamma^2 p^{-c_0}).
\end{align}
Combining (\ref{e046}) with (\ref{e113}) on the event $\mathscr{E}$ yields
$ E \|\bdelta\|_2^2 \leq   63^2\lambda_0^2s/\kappa^4(s,7)+ (3/4) \lambda_0^{-2} E\{\veps_0^4 1_{\{|\veps_0| > \tau\}}\} + O(\gamma^2 p^{-c_0})$.
For the $L_q$-loss with $q \in (1, 2)$, an application of H\"{o}lder's inequality and Young's inequality with (\ref{e044}) and (\ref{e046}) gives
\begin{align} \label{e073}
\nonumber
E \{\|\bdelta\|_q^q 1_{\mathscr{E}^c}\} & = E \big(\sum\nolimits_{j = 1}^p |\delta_j|^{2-q} |\delta_j|^{2q - 2} 1_{\mathscr{E}^c}\big) \leq \{E (\|\bdelta\|_1 1_{\mathscr{E}^c})\}^{2 - q} \{E (\|\bdelta\|_2^2 1_{\mathscr{E}^c})\}^{q - 1} \\
\nonumber
& \leq (2 - q) E \{\|\bdelta\|_1 1_{\mathscr{E}^c}\} + (q - 1) E \{\|\bdelta\|_2^2 1_{\mathscr{E}^c}\} \leq (2 - q) (2 \lambda_0)^{-1} E\{\veps_0^2 1_{\{|\veps_0| > \tau\}}\} \\
& \quad + (q - 1) (3/4) \lambda_0^{-2} E\{\veps_0^4 1_{\{|\veps_0| > \tau\}}\} + O[\{(2 - q) \gamma + (q - 1) \gamma^2\} p^{-c_0}],
\end{align}
where $\bdelta = (\delta_1, \ldots, \delta_p)\t$. It follows from this inequality and (\ref{e112}) on the event $\mathscr{E}$ that
\begin{align*}
E (\|\bdelta\|_q^q) & \leq 63^q \lambda_0^q s \kappa^{-2 q}(s,7) + (2 - q) (2 \lambda_0)^{-1} E\{\veps_0^2 1_{\{|\veps_0| > \tau\}}\} + (q - 1) (3/4) \lambda_0^{-2} E\{\veps_0^4 1_{\{|\veps_0| > \tau\}}\}  \\
& \quad+ O[\{(2 - q) \gamma + (q - 1) \gamma^2\} p^{-c_0}],
\end{align*}
which completes the proof for the first part of Theorem \ref{Thm2}.

The second part of Theorem \ref{Thm2} can be proved by noting $\sgn(\hbbeta) = \sgn(\bbeta_0)$ under the additional condition and using similar arguments as above.


\newpage

\setcounter{page}{1}
\setcounter{section}{0}
\setcounter{equation}{0}

\renewcommand{\theequation}{B.\arabic{equation}}
\setcounter{equation}{0}

\quad \bigskip

\begin{center}{\bf \Large Supplementary material for ``Asymptotic properties for\\ combined $L_1$ and concave regularization''}

\bigskip

\textsc{\large Yingying Fan and Jinchi Lv}
\bigskip
\end{center}

\noindent This Supplementary Material contains the proofs of Proposition \ref{Prop1} and Theorem \ref{Thm3}, and further details for \S\ref{Sec5}.

\section{Proof of Proposition \ref{Prop1}} \label{SecA.1}
Let $\hbbeta$ be any local minimizer of (\ref{e006}) that is the global minimizer along each coordinate. When constrained on one coordinate, the minimization problem (\ref{e006}) becomes a univariate penalized least-squares problem of form as in (\ref{e010}), with $p_{\text{H}, \lambda}(|\beta|)$ replaced by $p_\lambda(|\beta|)$, in which the value of scalar $z$ depends on the coordinate. When $p_\lambda(t)$ is chosen as the hard-thresholding penalty $p_{\text{H}, \lambda}(t)$, the univariate solution $\hbeta(z)$ is the soft-hard-thresholded estimator $\hbeta_{\text{SH}}(z)$ given in (\ref{e013}). Consider two cases.

\textit{Case 1}: $|z| \leq \lambda + \lambda_0$. Then $\hbeta_{\text{SH}}(z) = 0$, meaning that 0 is the global minimizer of $Q_H(\beta) = 2^{-1} (z - \beta)^2 + \lambda_0 |\beta| + p_{\text{H}, \lambda}(|\beta|)$. Denote by $Q(\beta)$ the function $Q_H(\beta)$ with $p_\lambda(|\beta|)$ in place of $p_{\text{H}, \lambda}(|\beta|)$. By assumption and $p_\lambda(t) \geq p_\lambda(\lambda) \geq p_{\text{H}, \lambda}(\lambda) = p_{\text{H}, \lambda}(t)$ for $t \geq \lambda$, it follows that $Q(\beta)\geq Q_H(\beta)$. These along with $Q(0) = Q_H(0)$ entail $\hbeta(z) = 0$.

\textit{Case 2}: $|z| > \lambda + \lambda_0$. Then by the monotonicity of $p_\lambda(t)$, $\hbeta(z)$ has the same sign as $z$. It follows from $p_\lambda'\{(1 - c_1) \lambda\} \leq c_1 \lambda$ and $|z| > \lambda + \lambda_0$ that
\[
\sgn(z) Q'\{\sgn(z) (1 - c_1) \lambda\} = (1 - c_1) \lambda - |z| + \lambda_0 + p_\lambda'\{(1 - c_1) \lambda\} \leq \lambda + \lambda_0 - |z|  < 0.
\]
Since $-p''_\lambda(t)$ is decreasing on $[0, (1 - c_1) \lambda]$, the function $Q(\beta)$ is convex, or first concave and then convex as $\beta$ varies from $0$ to $\sgn(z) (1 - c_1) \lambda$, in view of $Q''(\beta) = 1 + p''_\lambda(|\beta|)$. This shape constraint of $Q(\beta)$ between $0$ to $\sgn(z) (1 - c_1) \lambda$ along with $\sgn(z) Q'\{\sgn(z) (1 - c_1) \lambda\} < 0$ entails that its minimum on this interval is attained at $0$ or $\sgn(z) (1 - c_1) \lambda$. Since  $\sgn(z) Q'\{\sgn(z) (1 - c_1) \lambda\} < 0$, the point $\sgn(z) (1 - c_1) \lambda$ cannot be the global minimizer of $Q(\beta)$. Thus $\hbeta(z) = 0$ or $|\hbeta(z)| > (1 - c_1) \lambda$, which completes the proof.

\section{Proof of Theorem \ref{Thm3}} \label{SecA.4}
We first make a simple observation. Let $\hbbeta_\text{global}$ be the global minimizer of (\ref{e006}) with the $L_\infty$-constraint $\|n^{-1} \bX\t (\by - \bX \bbeta)\|_\infty \leq \lambda_0/2$. Then conditional on the event $\mathscr{E} = \{\|n^{-1} \bX\t \bveps\|_\infty \leq \lambda_0/2\}$, $\bbeta_0$ is a feasible solution to this new minimization problem. Thus the proof of Theorem \ref{Thm1} applies to show that $\hbbeta_\text{global}$ has the same asymptotic properties as in Theorem \ref{Thm1}. Let $\hbbeta$ be a computable local minimizer of (\ref{e006}) that is global minimizer in each coordinate produced by any algorithm satisfying $\|\hbbeta\|_0 \leq c_2 s$ and $\|n^{-1} \bX\t (\by - \bX \hbbeta)\|_\infty = O(\lambda_0)$. Since $\lambda \geq c_3 \lambda_0$, it follows from Theorem \ref{Thm1} that $\text{FS}(\hbbeta_\text{global}) = O(s)$, which entails that $\|\hbbeta_\text{global}\|_0 \leq \|\bbeta_0\|_0 + \text{FS}(\hbbeta_\text{global}) = O(s)$. Denote by $A = \supp(\hbbeta) \cup \supp(\hbbeta_\text{global})$. Then in view of $\|\hbbeta\|_0 \leq c_2 s$, we have $|A| \leq \|\hbbeta\|_0 + \|\hbbeta_\text{global}\|_0 = O(s) \leq c_4 s$ for some sufficiently large positive constant $c_4$. In light of $\|n^{-1} \bX\t (\by - \bX \hbbeta_\text{global})\|_\infty \leq \lambda_0/2$ and $\|n^{-1} \bX\t (\by - \bX \hbbeta)\|_\infty = O(\lambda_0)$, similarly as in the proof of Theorem 5 in Zhang \& Zhang (2012) we can show that
\begin{equation} \label{n001}
\|n^{-1} \bX_A\t \bX_A \bdelta_A\|_2 \leq |A|^{1/2} \|n^{-1} \bX_A\t \bX_A \bdelta_A\|_\infty = O(s^{1/2}) O(\lambda_0) = O(\lambda_0 s^{1/2}),
\end{equation}
where $\bX_A$ denotes a submatrix of $\bX$ consisting of columns in $A$ and $\bdelta_A$ denotes a subvector of $\bdelta = \hbbeta - \hbbeta_\text{global}$ consisting of components in $A$. Since by assumption $\min_{\|\bgamma\|_2 = 1,\ \|\bgamma\|_0 \leq c_4 s}n^{-1/2} \|\bX \bgamma\|_2 \geq \kappa_0$, the smallest singular value of $n^{-1/2} \bX_A$ is bounded from below by $\kappa_0$, which along with inequality (\ref{n001}) yields the same asymptotic bounds on $\|\bdelta\|_q$ and $n^{-1/2}\|\bX \bdelta\|_2$ as in Theorem \ref{Thm1}. Combining these results with Theorem \ref{Thm1} leads to desired asymptotic bounds on the $L_q$-estimation and prediction losses for the computable solution $\hbbeta$.  Since $\hbbeta$ is global minimizer in each coordinate, it follows from Proposition \ref{Prop1} that each component of $\hbbeta$ is either zero or of magnitude larger than $(1 - c_1) \lambda$. Thus we can obtain similar bound on $\mbox{FS}(\hbbeta)$ using the bound on $\|\hbbeta - \bbeta_0\|_2$ and the same argument as in Theorem \ref{Thm1}, which concludes the proof.

\section{Further details for \S\ref{Sec5}} \label{SecA.5}

\begin{table}[hb]
\def~{\hphantom{0}}
\tbl{Selection probabilities of most frequently selected genes with number up to median model size over 50 random splittings of lung cancer data}{%
\begin{tabular}{rccccrcccc}
 \\
ID & Lasso & $L_1$+SCAD & $L_1$+Hard & $L_1$+SICA & ID & Lasso & $L_1$+SCAD & $L_1$+Hard & $L_1$+SICA \\[5pt]
    4     & 0$\cdot$92  & 0$\cdot$88  & 0$\cdot$84  & 0$\cdot$76  & 7046  & 0$\cdot$28  & ---   & ---   & --- \\
    1271  & 0$\cdot$64  & 0$\cdot$48  & 0$\cdot$50   & 0$\cdot$38  & 7327  & 0$\cdot$36  & ---   & ---   & 0$\cdot$34 \\
    2421  & 0$\cdot$78  & 0$\cdot$72  & 0$\cdot$72  & 0$\cdot$60   & 8537  & 0$\cdot$46  & ---   & ---   & --- \\
    3250  & 0$\cdot$64  & 0$\cdot$38  & 0$\cdot$36  & 0$\cdot$34  & 9019  & 0$\cdot$30   & ---   & ---   & --- \\
    3381  & 0$\cdot$46  & 0$\cdot$44  & 0$\cdot$38  & 0$\cdot$30   & 9365  & 0$\cdot$28  & ---   & ---   & --- \\
    3508  & 0$\cdot$44  & 0$\cdot$24  & 0$\cdot$30   & 0$\cdot$30   & 9824  & 0$\cdot$32  & ---   & ---   & --- \\
    5229  & 0$\cdot$32  & ---   & ---   & ---   & 10386 & 0$\cdot$52  & 0$\cdot$38  & 0$\cdot$34  & 0$\cdot$40 \\
    5301  & 0$\cdot$84  & 0$\cdot$60   & 0$\cdot$66  & 0$\cdot$56  & 11957 & 0$\cdot$96  & 0$\cdot$94  & 0$\cdot$94  & 0$\cdot$94 \\
    5793  & 0$\cdot$64  & 0$\cdot$44  & 0$\cdot$42  & 0$\cdot$58  & 12114 & 0$\cdot$34  & ---   & ---   & --- \\
    6600  & 0$\cdot$60   & 0$\cdot$30   & 0$\cdot$34  & 0$\cdot$38  &       &       &       &       &
\end{tabular}}
\label{tab2}
\begin{tabnote}
$L_1$+SCAD, combined $L_1$ and smoothly clipped absolute deviation; $L_1$+Hard, combined $L_1$ and hard-thresholding; $L_1$+SICA, combined $L_1$ and smooth integration of counting and absolute deviation.
\end{tabnote}
\end{table}

\end{document}